%% file: BayesianApproach.tex
\documentclass[conference]{IEEEtran}

\usepackage{amsfonts}
\usepackage{amsmath}
\usepackage[]{changepage} 
\usepackage{float}
\usepackage[T1]{fontenc} 
\usepackage[]{graphicx}
\usepackage[utf8]{inputenc}
\usepackage{makecell}
\usepackage{multirow}
\usepackage{pbox}
\usepackage{relsize}
\usepackage[group-separator={,}]{siunitx}
\usepackage{soul}
\usepackage{tabularx}
\usepackage{tabu}
\usepackage[normalem]{ulem}
\usepackage[table]{xcolor}

\graphicspath{{./figures/}}

\newcommand{\Beta}{B}

\newcommand{\mathA}{\mathcal{A}}
\newcommand{\mathB}{\mathcal{B}}

\newcommand{\mathG}{\mathcal{G}}
\newcommand{\mathP}{\mathcal{P}}

\newcommand{\TP}{\operatorname{TP}}

\newcommand{\FP}{\operatorname{FP}}
\newcommand{\TPR}{\operatorname{TPR}}
\newcommand{\FPR}{\operatorname{FPR}}
\newcommand{\AUC}{\operatorname{AUC}}

\begin{document}

\title{A Bayesian Approach to Income Inference in a Communication Network}

\author{%
\IEEEauthorblockN{Martin Fixman\IEEEauthorrefmark{1}\IEEEauthorrefmark{2},
Ariel Berenstein\IEEEauthorrefmark{1},
Jorge Brea\IEEEauthorrefmark{1},
Martin Minnoni\IEEEauthorrefmark{1},
Matias Travizano\IEEEauthorrefmark{1},
Carlos Sarraute\IEEEauthorrefmark{1}
}
\IEEEauthorblockA{\IEEEauthorrefmark{1}Grandata Labs, Bartolome Cruz 1818, Vicente Lopez, Argentina}
\IEEEauthorblockA{\IEEEauthorrefmark{2}Universidad de Buenos Aires, Argentina}
\IEEEauthorblockA{\{mfixman, ariel, jorge, martin, mat, charles\}@grandata.com}
}

\maketitle

\begin{abstract}

The explosion of mobile phone communications in the last years occurs at a moment where data processing power increases exponentially.  Thanks to those two changes in a global scale, the road has been opened to use mobile phone communications to generate inferences and characterizations of mobile phone users.
In this work, we use the communication network, enriched by a set of users' attributes, to gain a better understanding of the demographic features of a population. Namely, we use call detail records and banking information to infer the income of each person in the graph.

\end{abstract}

\input{introduction}
\input{data_source}

\input{inference_methodology}

\input{results}
\input{multiple_categories}

\input{conclusion}

\bibliography{bibliography/sna}{}

\end{document}

%% file: introduction.tex

\section{Introduction}

In recent years, we have witnessed an exponential growth in the capacity to gather, store and manipulate massive amounts of data across a broad spectrum of disciplines: in astrophysics our capacity to gather and analyze massive datasets from astronomical observations has significantly transformed our capacity to model the dynamics of our cosmos; in sociology our capacity to track and study traits from individuals within a population of millions is allowing us to create social models at multiple scales, tracking individual and collective behavior both in space and time, with a granularity not even imagined twenty years ago.

In particular, mobile phone datasets provide a very rich view into the social interactions and the physical movements of large segments of a population. The voice calls and text messages exchanged between people, together with the call locations (recorded through cell tower usages), allow us to construct a rich social graph which can give us interesting insights on the users' social fabric, detailing not only particular social relationships and traits, but also regular patterns of behavior both in space and time, such as their daily and weekly mobility patterns~\cite{gonzalez2008understanding,ponieman2013human,sarraute2015city}.

Demographic factors play an important role in the constitution and preservation of social links. In particular concerning their age, individuals have a tendency to
establish links with others of similar age. This phenomenon is called  age homophily~\cite{mcpherson2001birds}, and has been verified in mobile phone communications graph~\cite{blumenstock2010mobile,sarraute2014} as well as the Facebook graph~\cite{ugander2011anatomy}.

Economic factors are also believed to have a determining role in both the social network's structure and dynamics. However, there are still very few large-scale quantitative analyses on the interplay between economic status of individuals and their social network. In~\cite{leo2015socioeconomic}, the authors analyze the correlations between mobile phone data and banking transaction information, revealing the existence of social stratification. They also show the presence of socioeconomic homophily among the networks participants using users' income, purchasing power and debt as indicators.

In this work, we leverage the socioeconomic homophily present in the cellular phone network to generate inferences of socioeconomic status in the communication graph. To this aim we will use the following data sources: (i) the Call Detail Records (CDRs) from an operator allow us to construct a social graph and to establish social affinities among users; (ii) banking reported income for a subset of their clients obtained from a large bank data source. We then construct an inferential algorithm that allows us to predict the socioeconomic status of users close to those for which we have banking information. To our knowledge, this is the first time both mobile phone and banking information has been integrated in this way to make inferences based on a social telecommunication graph.

%% file: data_source.tex

\section{Data Sources}\label{data_sources}

\subsection{Mobile Phone Data Source}

The data used in this study consist of a set \( \mathP \) of \textit{Call Detail Records} (CDRs), composed of voice calls and text messages from a telecommunication company (\textit{telco}) for a 3 month period.

Every CDR \( p \in \mathP \) contains the phone numbers of the caller and callee \( \left< p_o, p_d \right> \), which are anonymized using a cryptographic hash function for privacy reasons, the starting time \( p_t \), and, in the case of voice calls, the call duration \( p_s \). The latitude and longitude of the antenna used for each call \( \left< p_y, p_x \right> \), are also given for a subset of the data.

Given that our collection \( \mathP \) of CDRs are coming from one telephone company, we are able to reconstruct all communication links between clients of this company, as well as communications between the clients and other users, but we have no information on communications where neither users are clients of the telco company.

If we define \( N \) as the number of users of the telco, and \( \mathP_N \subseteq \mathP \) as the calls where \( \left( \forall p \in \mathP_N \right) p_o \in N \wedge p_d \in N \), we create a communications graph \( \mathG_N \) which contains only the users from the telco, and all the calls exchanged between them.

\subsection{Banking Information}

For this study we also used account balances for over 10 million clients of a bank  for a period of 6 months, denoted \( \mathB \). The data for each client \( b \in \mathB \) contains his phone number \( b_p \), anonymized with the same hash function used in \( \mathP \), and the reported income of this person over 6 months \( b_{s_0}, \ldots, b_{s_5} \). We average these 6 values to get \( b_s \), an estimate of a user's income.

The bank also provided us with demographic information for a subset of its clients \( \mathA \subseteq \mathB \). For each user \( u \in \mathA \) we are given the age \( u_a \) of the user, which allows us to observe differences in the income distribution according to the age. In another line of work, homophily with respect to age has been observed and used to generate inferences~\cite{brea2014}.

\begin{figure}[h]
\begin{center}
\includegraphics[width=0.98\columnwidth]
{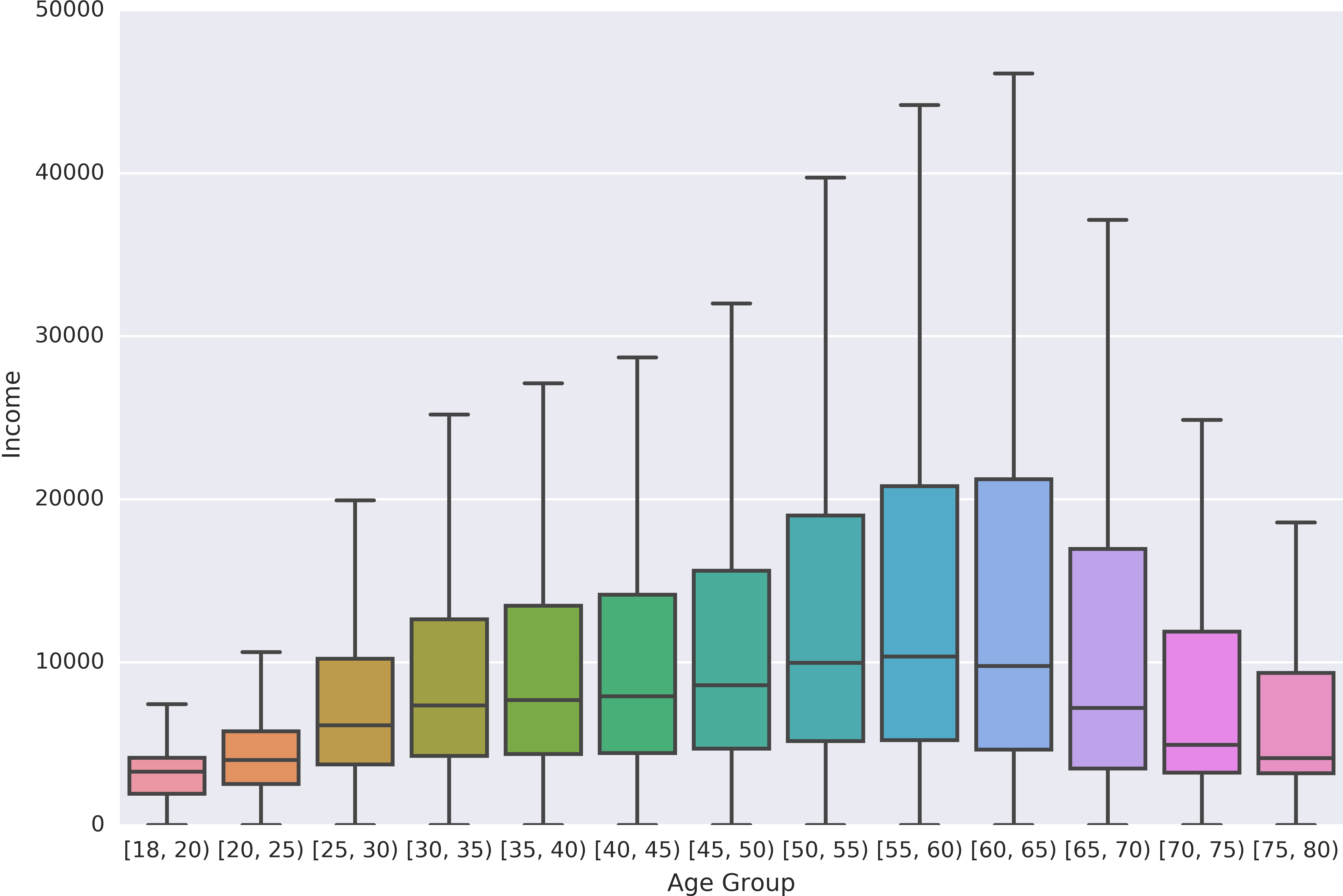}
\caption{Distribution of income $b_s$ as a factor of age $b_a$. This is consistent with data from median house income in the studied country~\cite{gallup2013}.}
\label{income_age_boxplot}
\end{center}
\end{figure}

Figure~\ref{income_age_boxplot} shows the distribution of income, according to the age range (generated by taking 5 years intervals for the age).
It is interesting to note how the median income increases with the age, up to
the 60--65 years range (the retirement age). After 65 years old, the median income rapidly decreases.

\subsection{Bank and Telco Matching}

Since the phone numbers in each call $ p_o $ and $ p_d $ are anonymized with the same hash function as the phone number in the bank data, $ b_p $, we can match users to their unique phone to create the social graph:
$$ G = \mathP \bowtie_{_{p_o = b_p}} \mathB \bowtie_{_{p_d = b_p}} \mathB $$
where $\bowtie$ denotes the inner join operator.
$G$ includes income information for the subset of the social graph that appears in the bank data, so \( \forall g \in G \) we have its phone number \( g_p \),  its average income over 6 months \( g_s \), and its age \( g_a \).
This graph has a total of \num{2027554} nodes with \num{5044976} edges, which represent \num{29599762} calls and \num{5476783} text messages.

\subsection{Outlier Filtering}

The dataset contains information about bank and telco users, some of which may not directly correspond to a human user, 
or may not have useful information for our research.
Most of the telco users in the first case are already filtered by the intersection (\textsc{inner join}). To make sure the users are relevant enough for this study, we only keep the users which have:

\begin{itemize}
	\item More than 5 calls in either direction.
	\item A monthly income of at least \$\num{54} (converted to US dollars).
	\item A monthly income in the \num{99}th percentile (i.e.\ we filter users with a monthly income in the top 1\%).
\end{itemize}

%% file: inference_methodology.tex

\section{Inference Methodology}
\label{inference_methodology}

\subsection{Income Homophily}

\begin{figure}[h]
\begin{center}
\includegraphics[width=\columnwidth]
{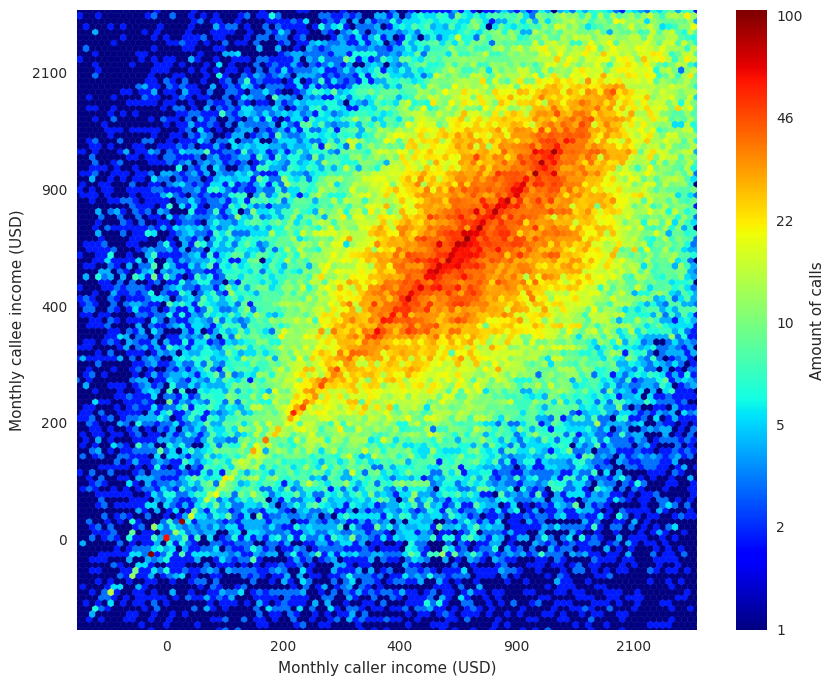}
\caption{Heatmap showing the number of calls between users, according to their monthly income. There is a higher probability that the callee and the caller have similar income levels.}
\label{homophily_heatmap}
\end{center}
\end{figure}

The main contribution of this work is the estimation of the income of the telco users for which we lack banking data, but have bank clients in their neighborhood of the network graph. To show the feasibility of this task, we first show the existence of a strong income homophily in the telco graph as is evidenced in Figure~\ref{homophily_heatmap}.

For each pair \( \left< o, d \right> \in \mathlarger{G} \), we define \( X \) as the set of incomes for callers and \( Y \) as the set of incomes for callees. According to what we can observe in Figure~\ref{homophily_heatmap}, \( X \) and \( Y \) should be significantly correlated. Given the broad non Gaussian distribution of the income's values, we choose to use a rank-based measure of correlation which is robust to outliers.
Namely we computed the \textit{Spearman's rank correlation} 
to test the statistical dependence of sets \( X \) and \( Y \):
\begin{equation}
r_s = \mathlarger{\rho}_{\operatorname{rank}(X) \operatorname{rank}(Y)} = \frac{\operatorname{cov}(\operatorname{rank}(x), \operatorname{rank}(y))}{\sigma_{\operatorname{rank}(X)} \sigma_{\operatorname{rank}(Y)}}
\label{spearman}
\end{equation}
this coefficient gives us a correlation coefficient of $\mathbf{r_s = 0.474}$. We also compared our result with a randomized null hypothesis, where links between users are selected randomly disregarding income data, obtaining a p-value of $ p < 10^{-6} $. These values for $r_s$ and $p$ show a strong indication of income homophily among users in our communication graph.
This observation is consistent with the results in~\cite{leo2015socioeconomic}.

We can take advantage of this homophily to propagate income information to the rest of our graph $ \mathP $, where we don't know the income of all of the users.

\subsection{Prediction Algorithm}

Instead of predicting the exact value of a user's income, our strategy is to distinguish between only two income categories depending on their monthly income (expressed in US dollars): $R_1 = \left[54, 340\right)$ and $R_2 = \left[340, \infty\right)$, that is, users with low or high income respectively, which we place into two distinct groups $ H_1, H_2 \subseteq G$ depending on \( g_s \), the users' income:
\[
	g \in H_i \iff g_s \in R_i
\]

We define the set $Q$ as the group of users having at least one connection link to bank clients. For each user $q^j \in Q$, we compute the number of outgoing calls $a^j_i$ to the category $H_i$. Our hypothesis, given the observed homophily, is that if a user $q^j$ has a higher number of calls $a^j_i$ to the category $H_i$ than the other category, it would be more likely to belong to the $H_i$ income category. In other words, a person is usually in the same income category as the majority of people it calls.

A straight forward approach would be to define the income category of a user as the category where most of its contacts belong. The problem with this approach is that it does not factor in the higher uncertainty in our estimates for users with fewer calls. To address this uncertainty, instead of using calling frequencies to define the probability of a user belonging to the high income category, we use the amount of calls $a^j_i$  as parameters defining a Beta distribution for the probability of belonging to a given category. We have therefore taken a Bayesian rather than a frequentist approach to income prediction.

We define \(\Beta^j\) as the Beta probability distribution function for each user.

\begin{equation}
	\Beta^j \left( x; \alpha^j, \beta^j \right) = \frac{1}{\Beta \left( \alpha^j, \beta^j \right)} x^{\alpha^j - 1} \cdot {\left( 1 - x \right)}^{\beta^j - 1}
\label{Beta}
\end{equation}
where $\alpha^j = a^j_1 +1$ and $\beta^j = a^j_2 +1$ are the parameters of the Beta distribution,
and $\Beta$ is the beta function, defined as:
\begin{equation}
\Beta \left( \alpha, \beta \right) =
\frac{\Gamma \! \left( \alpha \right) \cdot \Gamma \! \left( \beta \right)}
{\Gamma \! \left( \alpha + \beta \right)}
\label{Beta}
\end{equation}

Note that the above equation defines a distinct distribution for each user. Having obtained the Beta distribution for the probability of belonging to the high income category, we then find the lowest 5 percentile $p_{lower}$ for this probability. If $p_{lower}$ is above a given threshold $\tau$, we set the user's income to $H_2$, otherwise we set his income category to $H_1$. We note that this criteria takes into account both the mean and the broadness (uncertainty) of the distribution. We also note that the category assigned to a user depends not only on its Beta distribution but also on our choice of $\tau$.

%% file: results.tex

\section{Results}

We describe in this section the validation of our methodology.
We examine the true positive ($\TPR$) and false positive ($\FPR$) rates, \( \TPR = \TP / \operatorname{P} \) and \( \FPR = \FP / \operatorname{N} \), where $\TP$ is the number of correctly predicted users with high income, $\operatorname{P}$ is the total number of users with high income, $\FP$ is the number of users incorrectly classified as having high income, and $\operatorname{N}$ is the total number of users with low income.

\begin{figure}[H]
\begin{center}
{\includegraphics[width=0.95\columnwidth]
{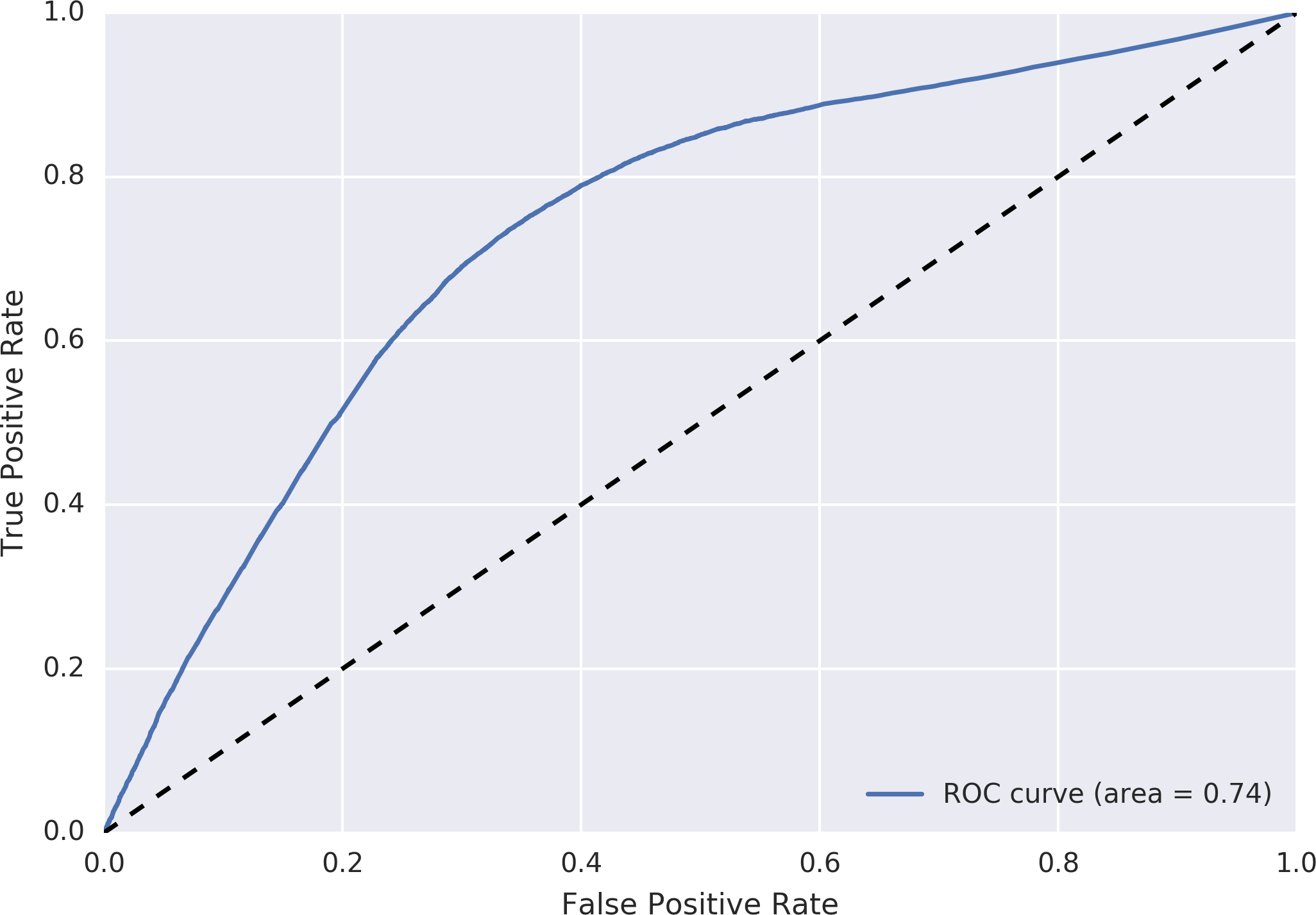}
}
\caption{ROC curve for prediction procedure. We observed an $\AUC = 0.74$ indicating that our predictor is better than a random predictor ($\AUC \simeq 0.50$).}
\label{ROC_multiclass}
\end{center}
\end{figure}

In Figure~\ref{ROC_multiclass} we plot the ROC (\textit{Receiver Operating Characteristic}) curve, showing $\TPR$ and $\FPR$ for the set of possible values of $\tau$. We see that our methodology clearly outperforms random guessing (dashed straight line). We can summarize our performance by calculating  the $\AUC$ (\textit{Area Under the Curve}) which in Figure~\ref{ROC_multiclass} is $\AUC = 0.74$. Note that random guessing would give a value of $\AUC \simeq 0.50$.

Alternatively we can evaluate the performance of our model by computing its accuracy for a given considered threshold $\tau$. The best accuracy obtained is \num{0.71} for $\tau = 0.4$. 

\subsection{Comparison with other inference methods}

We applied two other inference methods to the same data and compared their accuracies to our Bayesian model.

\begin{itemize}
	\item \textbf{Random selection} which chooses randomly the category for each user.
	\item \textbf{Majority voting} which decides whether a user is in the high or low income category depending on the category of the majority of its contacts. In case of a tie, the category is chosen randomly.
\end{itemize}

The accuracy of the first method is as expected \num{0.50}, while the accuracy for majority voting is \num{0.66}. 
With the Bayesian method we obtain an accuracy of \num{0.71}.

%% file: multiple_categories.tex

\section{Extension to Multiple Categories}

We present here how the methodology described in Section~\ref{inference_methodology} for
two categories can be extended to multiple categories.
To this end, we separate the income values into five distinct groups $ H_1, \ldots, H_5 \subseteq G$ of increasing wealth. A user is part of this group if its income is between the defined bounds, that is, \( g \in H_i \iff g_s \in R_i \).

The income ranges are set as follows:
	$R_1 = \left[54, 135\right) $;
	$R_2 = \left[135, 405\right) $;
	$R_3 = \left[405, 1080\right) $;
	$R_4 = \left[1080, 2700\right) $;
	$R_5 = \left[2700, \infty\right) $.

Again, we define the set $Q$ as the group of users having at least one connection link to bank clients. For each user $q^j \in Q$, we compute the number of outgoing calls $a^j_i$ to the category $H_i$.
We use the amount of calls $a^j_i$  as parameters defining a Dirichlet distribution for the probability of belonging to each category.
We define below the Dirichlet probability distribution function $D^j$:
\begin{equation}
D^j \left( x_1, \ldots, x_5; \alpha^j_1, \ldots, \alpha^j_5 \right) = \frac{1}{\Beta \left( \alpha \right)} \prod^5_{i = 1} x_i^{\alpha^j_i - 1}
\label{Dirichlet}
\end{equation}
where $\alpha^j_i = a^j_i +1$ are the parameters of the Dirichlet distribution, and $\Beta$ is the multivariate beta distribution function, defined by: 
\begin{equation}
\Beta \left( \alpha_1, \ldots, \alpha_k \right) = \frac{\prod^k_{i = 1} \Gamma \! \left( \alpha_i \right)}{\Gamma \! \left( \sum^k_{i = 1} \alpha_i \right) }
\label{Beta}
\end{equation}

Note that the above equation defines a distinct Dirichlet distribution for each user. For each of these distributions, we computed the marginal probability functions across all different categories, which result in Beta distributed functions, and use them to get the lowest 5 percentiles (\(p^i_{\operatorname{lower}}\)) in each case ${i=1, \ldots, 5}$ which can be compared to assign a category to each user.

\begin{figure}
\centering
\includegraphics[width=0.90\columnwidth]{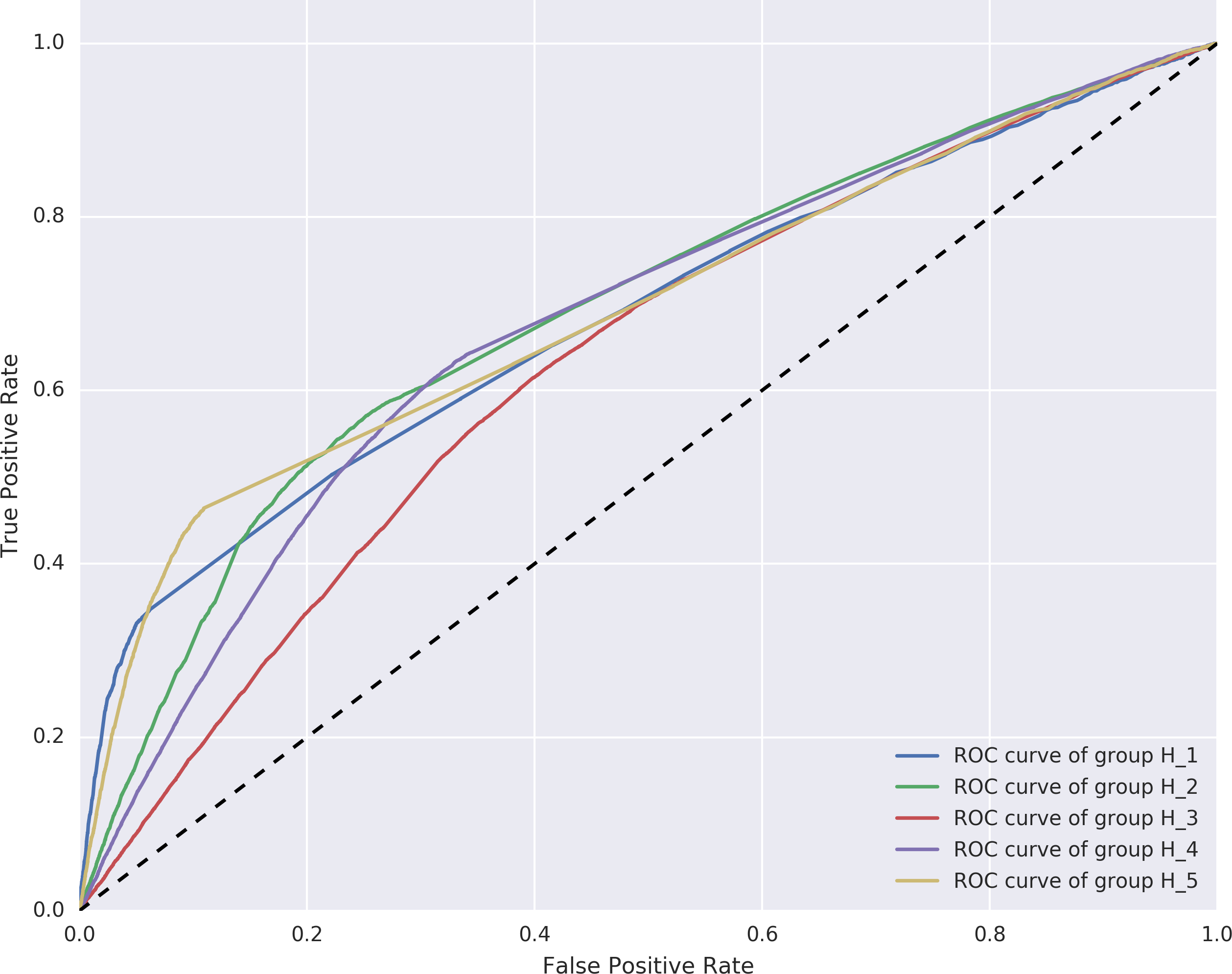}
\caption{ROC curves for multiclass problem. The performances observed are: $\AUC_1 = 0.68$, $\AUC_2 = 0.69$, $\AUC_3 = 0.63$, $\AUC_4 = 0.68$, $\AUC_5 = 0.69$. These predictors perform better than the random case, and have a similar performance (with exception of category 3).}
\label{roc_multiple_categories}
\end{figure}

In order to gain an intuition on how the classification extends to the multiple category case, we constructed for each category $i$ a binary classifier by using the computed $p^i_{lower}$ score and a given threshold $\tau$. In each case we sweep the threshold $\tau$ and compute the resulting ROC curves as shown in Figure~\ref{roc_multiple_categories}.

We observed the performance for the different categories: $\AUC_1 = 0.68$, $\AUC_2 = 0.69$, $\AUC_3=0.63$, $\AUC_4 = 0.68$, $\AUC_5 = 0.69$. In all cases, the predictor performs better than the random case.

%% file: conclusion.tex
\section{Conclusion}

This work is based on a combined data source
from mobile phone records and banking information.
We first showed the presence of homophily with respect to the monthly income in the
communication graph, that is, users of similar income tend to communicate more with each other. Based on the evidenced homophily, we presented a methodology to infer income categories for users in the graph for which we don't have
banking information, by taking a Bayesian approach.

To classify users into income categories, we first computed the number of calls a user $u$ makes to members of the different categories, and used them to construct a Beta distribution for the probability of user $u$ to belong to the high income category. We then validated our approach by constructing the ROC curve which clearly shows that our methodology outperforms random guessing.
Finally, we described how we can extend our approach to multiple categories using the Dirichlet distribution.

Our proposed inference methodology is useful for concrete applications, since it provides an estimation of socio-economic attributes of users lacking banking history, based on their communication network. We also note that this methodology is not restricted to the inference of socio-economic attributes, but is equally applicable to any attribute that exhibits significant homophily in the network.